\documentclass[wssquare]{ws-procs961x669}            

\usepackage{ws-toc,ws-multind}
\usepackage[dvipsnames]{xcolor}
\usepackage{orcidlink}
\hypersetup{colorlinks, citecolor=BrickRed}
\hypersetup{colorlinks=true, linkcolor=Blue, urlcolor=MidnightBlue}

\makeindex{author}                     

\begin{document}

\markboth{J. M. Z. Pretel, M. Dutra and S. B. Duarte}
{Neutron stars with a dark-energy core from the Chaplygin gas}

\title{ Neutron stars with a dark-energy core from the Chaplygin gas\footnote{Presented by Juan M. Z. Pretel at the XVII Marcel Grossmann Meeting, Pescara, Italy, 7-12 July 2024} }

\author{Juan M. Z. Pretel }
\address{Centro Brasileiro de Pesquisas F{\'i}sicas, Rua Dr. Xavier Sigaud, 150 URCA \\
Rio de Janeiro CEP 22290-180, RJ, Brazil \\
\textcolor{magenta}{juanzarate@cbpf.br}}

\author{Mariana Dutra}
\address{Departamento de F{\'i}sica, Instituto Tecnol{\'o}gico de Aeron{\'a}utica  \\
DCTA, 12228-900, S{\'a}o Jos{\'e} dos Campos, SP, Brazil  \\
\textcolor{magenta}{marianad@ita.br}}

\author{Sergio B. Duarte}
\address{Centro Brasileiro de Pesquisas F{\'i}sicas, Rua Dr. Xavier Sigaud, 150 URCA \\
Rio de Janeiro CEP 22290-180, RJ, Brazil \\
\textcolor{magenta}{sbd@cbpf.br}}

\begin{abstract}
We analyze the effect of a Chaplygin dark fluid (CDF) core on neutron stars (NSs). To address this study, we focus on the relativistic structure of stellar configurations composed by a dark-energy core, described by a Chaplygin-like equation of state (EoS), and an ordinary-matter crust which is described by a polytropic EoS. We examine the impact of the rate of energy densities at the discontinuous surface, defined as $\alpha= \rho_{\rm dis}^-/\rho_{\rm dis}^+$, on the radius, total gravitational mass, oscillation spectrum and tidal deformability. Furthermore, we compare our theoretical predictions with several observational mass-radius measurements and tidal deformability constraints. These comparisons together with the radial stability analysis show that the existence of NSs with a dark-energy core is possible. 
\end{abstract}

\bodymatter

\section{Introduction} 

Different cosmological observations have confirmed that the
Universe is experiencing an accelerated expansion era \cite{Riess1998, Perlmutter1999}. Since then, several dark energy (DE) models have been proposed to account for this late-time cosmic expansion \cite{Ratra1988, Kamenshchik2001, Copeland2006, Li2011, Joyce2016, Koyama2016, Montefalcone2020, Shankaranarayanan2022}. In the framework of Einstein gravity, the well-known $\Lambda$CDM model is based on cosmological constant $\Lambda$ and cold dark matter, where the latter is postulated in order to account for the gravitational effects observed in very large-scale structures. According to such a model, as the Universe continues to expand over time, the negative pressure associated with $\Lambda$ increasingly dominates over the attractive gravitational forces, and the expansion of the Universe accelerates. However, the $\Lambda$CDM model suffers from some problems that motivate the search for other phenomenological and theoretical models.

Among the available theoretical models, some researchers consider that the DE contribution might come in an EoS of the form $p= w\rho$, where $\rho$ and $p$ are the energy density and pressure of dark energy, respectively. From this perspective, the acceleration equation $\ddot{a}/a = -4\pi(\rho+3p)/3$ provides a positive acceleration $(\ddot{a}> 0)$ if $w <-1/3$ for a component with $\rho>0$. The special case $w=-1$ represents a cosmological constant, while quintessence models lie in the range $-1< w < -1/3$.  There are of course several quintessence models which predict different bounds for the EoS parameter $w$ \cite{Wang2000, Douspis2003}. Moreover, $w< -1$ for a ``phantom'' energy component \cite{Caldwell2002}. Within an astrophysical scenario, compact stars (composed of DE) have also been investigated by adopting the EoS $p= w\rho$ with negative $w$ \cite{Rahaman2012, Bhar2015, Banerjee2020}, which is one of the more promising directions to elucidate the  late-time accelerated expansion of the Universe. Nevertheless, the literature also provides other phenomenological models to describe the dark components of the Universe.

The exact physical nature of DE is still a mystery and, therefore, the possibility that dark matter and DE could be different manifestations of a single substance has been considered \cite{Kamenshchik2001, Bento2002, Reis2003, Xu2012}. In that regard, it has been shown that Chaplygin gas offers a simple unified model of dark matter and DE \cite{BILIC2002}. In other words, this model behaves like a cosmological constant at late stage and as dust-like matter at early stage. In Addition, in the light-cone parameterization, the original Chaplygin gas model can be obtained from the string Nambu–Goto action for $d$-branes moving in a $(d+2)$-dimensional spacetime \cite{Bordemann1993, Ogawa2000}. Some authors also argued that a cosmological constant would be ruled out if the Universe is dominated by a CDF \cite{Makler2003}. All these remarkable features of the Chaplygin gas motivate us to consider it when studying DE.

The effects of a DE fluid on the relativistic structure of single-phase compact stars (described by a Chaplygin-type EoS) have been intensively investigated in recent years \cite{Rahaman2010, Bhar2018, Panotopoulos2020EPJP, Tello2020, Panotopoulos2021, Prasad2021, Pretel2023EPJC, Jyothilakshmi2024}. Indeed, for such an EoS, it has been shown that the stellar structure equations provide maximum masses above $2\, M_\odot$ \cite{Panotopoulos2020EPJP, Panotopoulos2021, Pretel2023EPJC, Jyothilakshmi2024}, which favors the observational measurements. Furthermore, these stellar configurations made of a CDF obey the causality condition and are dynamically stable against radial pulsations when $dM/d\rho_c>0$ on the $M(\rho_c)$ curve \cite{Panotopoulos2020EPJP, Pretel2023EPJC}, where $\rho_c$ is the central energy density. The main purpose of the present study is to extend previous works to a hybrid context, where the DE is confined to the stellar core while the crust is ordinary matter described by a polytropic EoS. To examine the possible existence of such compact stars with a CDF core in the Universe, we will study their radial stability and compare their mass-radius relations as well as their tidal deformabilities with different astrophysical observations.

To achieve our results, the present work is organized as follows: In Sec.~\ref{Sec2} we present all the differential equations describing the different macroscopic properties of a compact star in general relativity. Our numerical results as well as their discussion are provided in Sec.~\ref{Sec3}, and finally we conclude in Sec.~\ref{Sec4}. In this paper we will use a geometric unit system and the sign convention $(-,+,+,+)$. However, our results will be given in physical units.


\section{Stellar structure equations}\label{Sec2} 

\subsection{TOV equations and equation of state}

In general relativity, the differential equations governing the hydrostatic equilibrium of a compact star are known as the Tolman-Oppenheimer-Volkoff (TOV) equations, namely,
\begin{align}
    \frac{dm}{dr} &= 4\pi r^2\rho ,  \label{TOV1}  \\
    \frac{dp}{dr} &= -(\rho+ p)\left[ \frac{m}{r^2}+ 4\pi rp \right]\left[ 1- \frac{2m}{r} \right]^{-1} ,  \label{TOV2}  \\
    \frac{d\psi}{dr} &= -\frac{1}{\rho+ p}\frac{dp}{dr} ,  \label{TOV3}
\end{align}
which are obtained from the Einstein field equations $G_{\mu\nu}= 8\pi T_{\mu\nu}$. The matter-energy content is described by an isotropic perfect fluid whose energy-momentum tensor is given by $T_{\mu\nu}= (\rho+ p)u_\mu u_\nu+ pg_{\mu\nu}$, where $\rho$ stands for the energy density, $p$ is the pressure and $u^\mu$ is the four-velocity. Furthermore, it is assumed that stellar configurations are spherically symmetric, i.e., described by the metric
\begin{equation}\label{metricEq}
    ds^2= -e^{2\psi}dt^2 + e^{2\lambda}dr^2 + r^2(d\theta^2+ \sin^2\theta d\phi^2) .
\end{equation}

The metric variable $\lambda(r)$ is determined from the relation $e^{-2\lambda}= 1- 2m/r$, where $m(r)$ is a mass function along the radial coordinate. Given an EoS of the form $p= p(\rho)$, the number of variables in the system of differential equations (\ref{TOV1})-(\ref{TOV3}) is reduced to three and therefore three boundary conditions are required:
\begin{align}\label{BCforTOV}
    \rho(0) &= \rho_c,  &  m(0) &= 0,  &  \psi(R) &= \frac{1}{2}\ln\left[ 1- \frac{2M}{R} \right] ,
\end{align}
where $\rho_c$ is the central energy density and will be varied within a certain interval in order to obtain a family of equilibrium configurations. The radius of the star $R$ is determined when the pressure drops to zero, so that the total gravitational mass is $M= m(r=R)$.

The functional relation between the energy density $\rho$ and pressure $p$ is known as EoS and is a crucial input when solving the TOV equations. In our hybrid stellar model we confine the DE to the core of the compact star, with the normal-matter crust surrounding it. The core is described by a Chaplygin-type EoS, while the ordinary matter in the outer layer or envelope is described by a polytropic EoS. Thus, the EoS for the two-phase stellar fluid is given by \cite{Pretel2024}
\begin{equation}\label{EoSEq}
    p(\rho) = 
  \begin{cases}
    A\rho- \frac{B}{\rho},  & \quad  0 \leq r \leq R_{\rm dis} ,  \\ 
    \kappa\rho^{1+ 1/\eta},  & \quad  R_{\rm dis} \leq r \leq R ,
  \end{cases}
\end{equation}
with $R_{\rm dis}$ being the radius of the discontinuous surface. The extra term ``$-B/\rho$'', where $B$ is a positive constant (given in $\rm m^{-4}$ units), indicates a negative pressure that leads to the accelerated expansion of the Universe \cite{Kamenshchik2001}. Meanwhile, the contribution ``$A\rho$'' describes a barotropic fluid, where the CDF parameter $A$ is a positive dimensionless constant. In addition, for the polytropic EoS, we will establish $\eta= 1.0$ and $\kappa= 100\, \rm km^2$, which are typical values to describe neutron stars \cite{KokkotasRuoff2001, Pretel2021JCAP}. Of course, the microscopic properties of NSs involve more realistic EoSs, but the use of a polytropic EoS as a first analysis in the form of a toy model is a reasonable consideration. More realistic EoS models describing the crust will be considered in future studies.

At the splitting wall, the pressure must be continuous, so we can write the CDF parameter $B$ as a function of the other model constants;
\begin{equation}\label{EqB}
    B = A(\rho_{\rm dis}^+)^2 - \kappa (\rho_{\rm dis}^+)(\rho_{\rm dis}^-)^{1+ 1/\eta} ,
\end{equation}
where $\alpha = \rho_{\rm dis}^-/\rho_{\rm dis}^+ \leq 1$ is the ratio of the outer density to the inner density at $r= R_{\rm dis}$. In other words, going from the center to the surface, $\rho_{\rm dis}^+$ is the energy density where the core ends, while $\rho_{\rm dis}^-$ is the energy density where the envelope of the hybrid star begins. For a fixed value of $\rho_{\rm dis}^+$, we see that small $\alpha$ leads to a larger jump in energy density. Given a central density value $\rho_c$, we will express our results in terms of the set of parameters $\{\rho_{\rm dis}^+, A, \alpha\}$.

\subsection{Radial pulsations}

The TOV equations provide equilibrium solutions, but such equilibrium may be stable or unstable with respect to small radial perturbations. A compact star is stable if its eigenfunctions $\omega_n^2$ are positive, where $n$ denotes the number of nodes between the center and the surface. Chandrasekhar pioneered the radial stability of single-phase relativistic stars \cite{ChandrasekharApJ, ChandrasekharPRL}, and since then the radial oscillation equations have been written in different forms \cite{VathChanmugam, Gondek1997, KokkotasRuoff2001, Pretel2020MNRAS, Bora2021, Hong2022}. For numerical convenience, in the present study we will use the differential equations obtained by Gondek \textit{et al}.~\cite{Gondek1997}. The linearized perturbation equations can be obtained by introducing the Lagrangian displacement $\xi$ around the equilibrium position, so that in the perturbed system we can write $\xi(t,r)= \chi(r)e^{i\omega t}$. Defining $\zeta= \chi/r$, the adiabatic radial pulsations of relativistic stars in Einstein gravity are governed by the following first-order time-independent equations
\begin{align}
    \dfrac{d\zeta}{dr} =& -\dfrac{1}{r}\left( 3\zeta + \dfrac{\Delta p}{\gamma p} \right) + \dfrac{d\psi}{dr}\zeta ,  \label{ROEq1} \\
    \dfrac{d(\Delta p)}{dr} = &\ \zeta\left[ \omega^2e^{2(\lambda - \psi)}(\rho + p)r - 4\dfrac{dp}{dr} - 8\pi e^{2\lambda}(\rho+ p)rp + r(\rho+ p)\left( \dfrac{d\psi}{dr} \right)^2  \right]   \nonumber  \\
    & - \Delta p\left[ \dfrac{d\psi}{dr} + 4\pi (\rho +p)re^{2\lambda} \right] ,  \label{ROEq2}
\end{align}
where $\gamma = (1 + \rho/p)dp/d\rho$ is the adiabatic index and $\Delta p$ denotes the Lagrangian perturbation of the pressure. We notice that Eq.~(\ref{ROEq1}) has a singularity at the stellar center ($r=0$), so that the condition $\Delta p = -3\zeta\gamma p$ guarantees regularity as we approach the center. On the other hand, at the surface ($r=R$), we must require $\Delta p = 0$.

However, our study is dealing with two-phase compact stars, and it is necessary to adopt suitable junction conditions at the discontinuous surface $r= R_{\rm dis}$. According to Pereira \textit{et al.}~\cite{Pereira2018}, these conditions depend on the velocity of the phase transition near the discontinuous wall, and are given by
\begin{itemize}
    \item[$\star$] For slow phase transitions, there is no mass transfer from one phase to another, and the junction conditions at the phase-splitting interface are given by
    \begin{align}\label{JuncCond1}
        \left[ \zeta \right]_-^+ &= 0,  &  \left[ \Delta p \right]_-^+ &= 0,
    \end{align}
    where $[z]_-^+ = z^+- z^-$, with $z$ standing for any variable across the splitting wall.

    \item[$\star$] For rapid phase transitions, there is a mass transfer between the two phases. The matching conditions at the interface are as follows
    \begin{align}\label{JuncCond2}
        \left[ \zeta- \frac{\Delta p}{rp'} \right]_-^+ &= 0,  &  \left[ \Delta p \right]_-^+ &= 0,
    \end{align}
    where $p' = dp/dr$ is defined in the hydrostatic equilibrium state.
\end{itemize}

\subsection{Tidal deformability}

In addition to mass-radius relations, the tidal deformation is another astrophysically observable macroscopic property of a NS. Indeed, these stars are tidally deformed under the presence of a companion star, and such a deformation can be inferred through the gravitational radiation emitted during the inspiral phase of compact binary systems \cite{Chatziioannou2020, Dietrich2021}. Since this quantity can be used to study stellar interiors, it also becomes important to investigate the effects of a CDF core on NSs. The tidal Love number $k_2$ is calculated by means of the expression
\begin{align}
k_2 &= \frac{8}{5}(1- 2C)^2C^5 \left[ 2C(y_R -1) - y_R+ 2 \right]  \nonumber  \\
&\times \left\lbrace 2C[ 4(y_R+ 1)C^4 + (6y_R- 4)C^3 \right.  \nonumber  \\
&\left.+\ (26- 22y_R)C^2 + 3(5y_R -8)C - 3y_R+ 6 \right]   \nonumber  \\
&\left.+\ 3(1-2C)^2\left[ 2C(y_R- 1)- y_R +2 \right]\ln(1-2C) \right\rbrace^{-1} ,
\label{LoveNumEq}
\end{align}
where $C= M/R$ is the compactness of the star of mass $M$ and radius $R$. Accordingly, the dimensionless tidal deformability is determined from $\Lambda= 2k_2C^{-5}/3$.

The surface value $y_R= y(R)$ is calculated after solving the differential equation
\begin{equation}\label{yEq}
    r\frac{dy}{dr} = -y^2 + (1 - r\mathcal{P})y - r^2\mathcal{Q} ,
\end{equation}
with the initial condition $y(0)=2$ \cite{Postnikov2010}, where 
\begin{align}
    \mathcal{P} =&\ \frac{2}{r} + e^{2\lambda}\left[ \frac{2m}{r^2} + 4\pi r(p - \rho) \right] ,  \\
    \mathcal{Q} =&\ 4\pi e^{2\lambda}\left[ 5\rho + 9p + \frac{\rho+ p}{dp/d\rho} \right] - \frac{6e^{2\lambda}}{r^2} - 4\psi'^2 .  \label{EqforQ}
\end{align}
Again we must keep in mind that our stellar model is a hybrid system, and Eq.~(\ref{yEq}) must be solved using a suitable junction condition at the interface. For first-order transitions in hybrid stars, it has been shown that the perturbation function $y(r)$ must satisfy the following junction condition at the splitting wall \cite{Postnikov2010, Janos2020}
\begin{equation}\label{BCTidalDef}
    [y]_-^+ = \frac{4\pi R_{\rm dis}^3(\rho_{\rm dis}^- - \rho_{\rm dis}^+)}{m(R_{\rm dis}) + 4\pi R_{\rm dis}^3 p(R_{\rm dis})} .
\end{equation}


\section{Numerical results and discussion}\label{Sec3}

Before starting to discuss our results we must emphasize that all differential equations describing the internal structure of a hybrid star must be solved separately, i.e., for the core and for the crust with the appropriate matching conditions already mentioned above. 

\begin{figure}[t]
\begin{center}
\includegraphics[width=12.6cm]{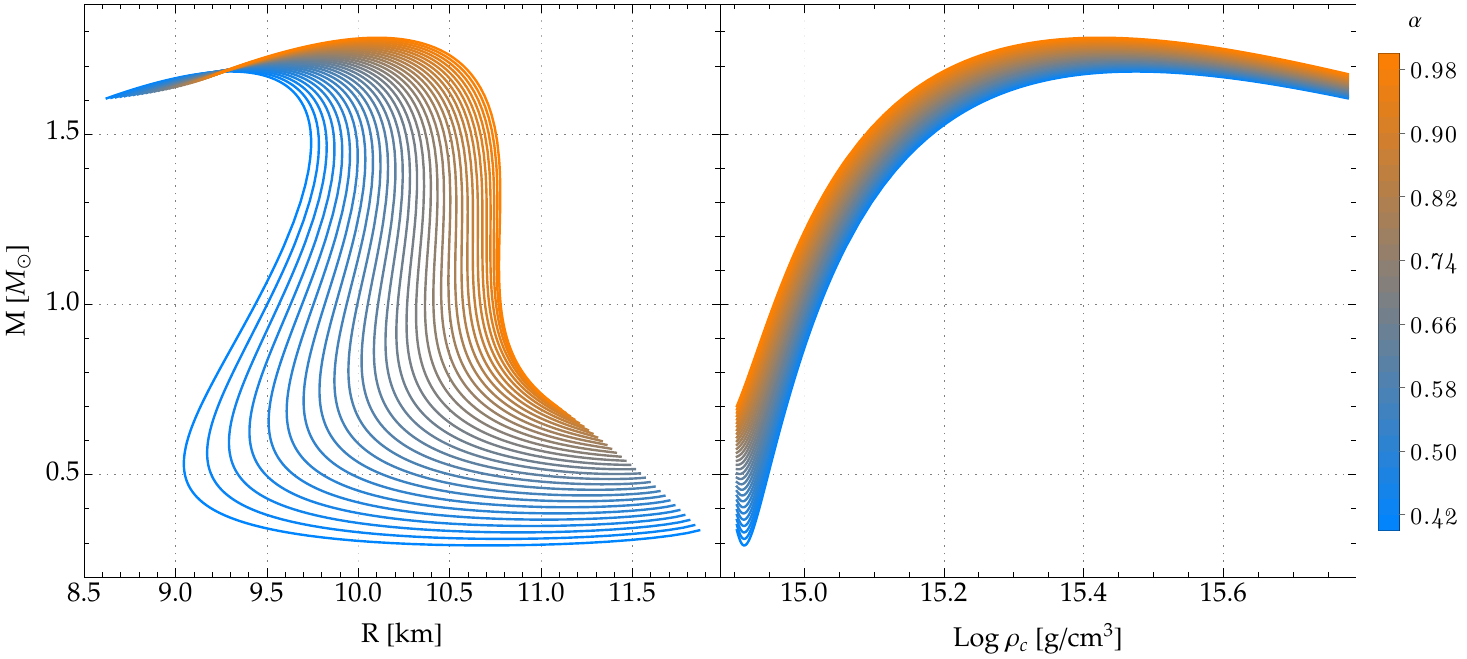}
\end{center}
\caption{Mass-radius profile (left panel) and mass-central density relation (right panel) for hybrid stellar models with EoS (\ref{EoSEq}) with inner energy density $\rho_{\rm dis}^+= 0.8 \times 10^{15}\, \rm g/cm^3$, CDF parameter $A= 0.3$ and a wide range of values of $\alpha \in [0.4, 1.0]$. According to the right plot, for sufficiently small values of $\alpha$ and lower central densities, we have $dM/d\rho_c <0$ which would indicate unstable compact stars.}
\label{FifMR1}
\end{figure}

\begin{figure}
\begin{center}
\includegraphics[width=12.6cm]{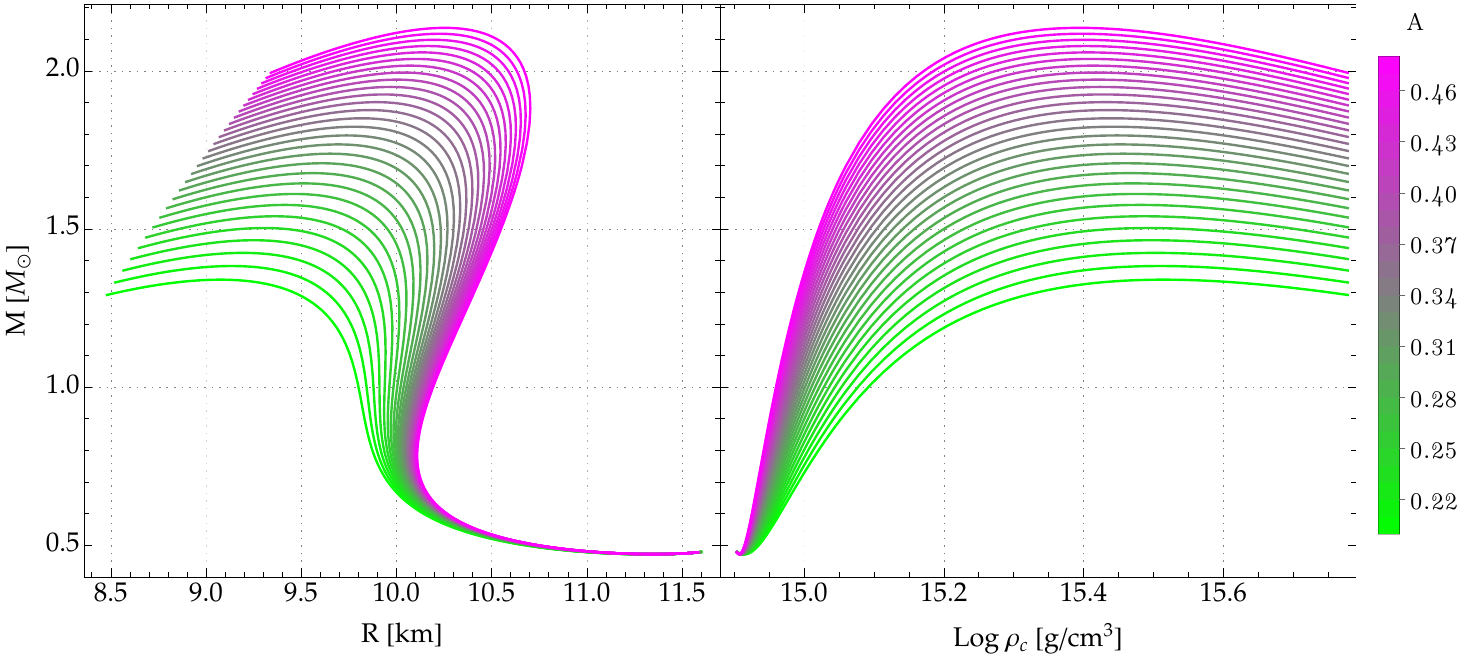}
\end{center}
\caption{Mass-radius diagram (left) and mass-central density relation (right) as in Fig.~\ref{FifMR1}, but for a fixed $\alpha= 0.6$ and CDF parameter varying in the interval $A\in [0.20, 0.48]$.}
\label{FifMR2}
\end{figure}

For $\rho_{\rm dis}^+= 0.8 \times 10^{15}\, \rm g/cm^3$ and CDF parameter $A= 0.3$, Fig.~\ref{FifMR1} displays the mass as a function of radius (left) and central density (right) for a wide range of values of $\alpha \in [0.4, 1.0]$. Given a $\rho_c$, it is observed that the total gravitational mass of the star increases with increasing $\alpha$. On the other hand, for a fixed $\alpha= 0.6$ and varying $A$ in the interval $A\in [0.20, 0.48]$, our results in Fig.~\ref{FifMR2} indicate that the increase in $A$ leads to an increase in the mass of the hybrid star. The impact of $A$ on the $M-R$ relation is substantial at high masses, but irrelevant in the low-mass region. Nevertheless, according to Fig.~\ref{FifMR1}, the largest effect of the parameter $\alpha$ on the $M-R$ diagram occurs at low masses.

\begin{figure}
\begin{center}
\includegraphics[width=6.2cm]{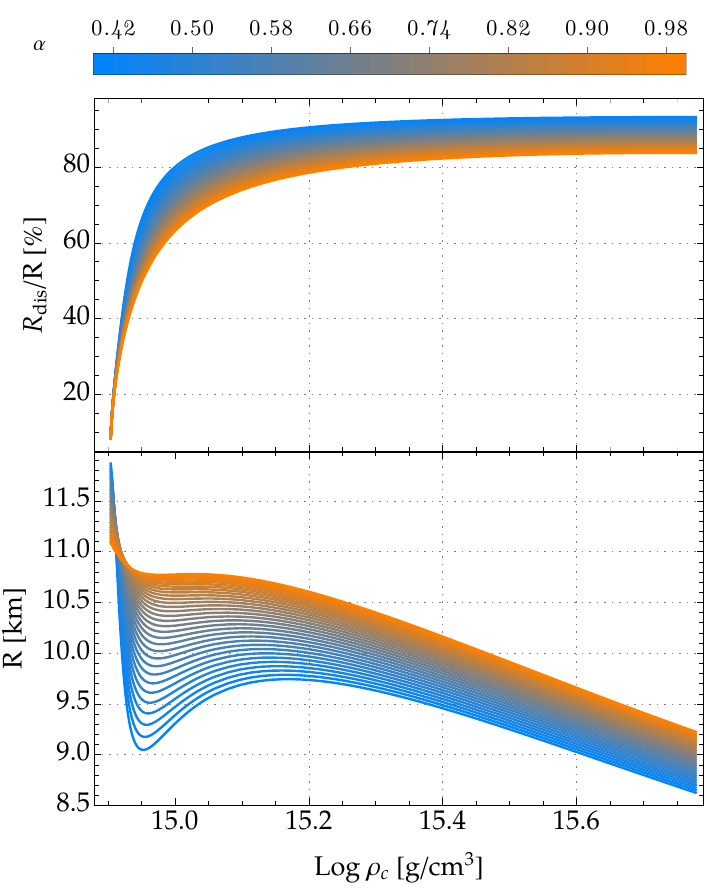}
\includegraphics[width=6.2cm]{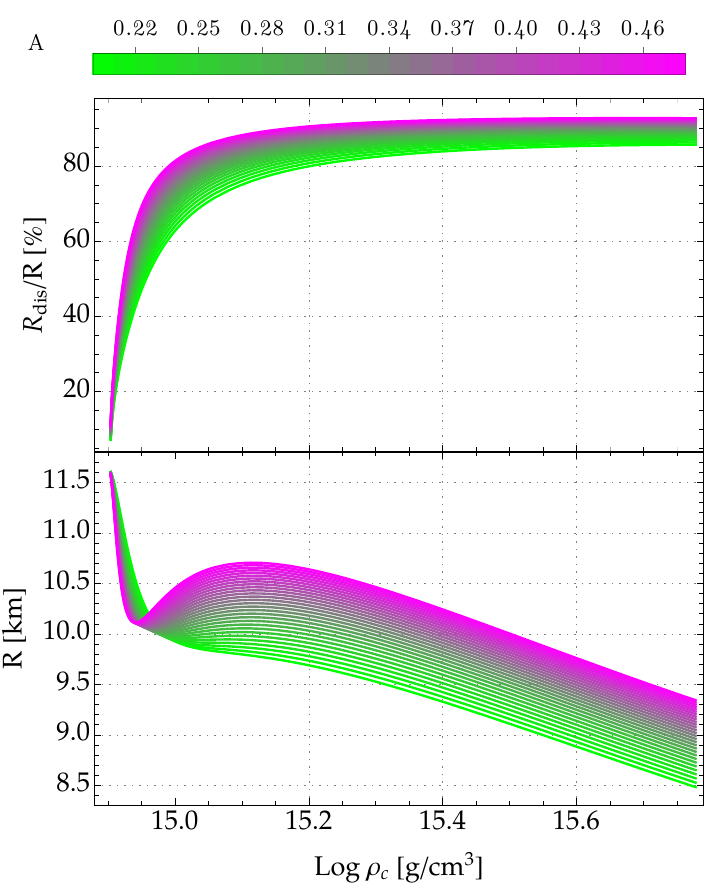}
\end{center}
\caption{Percentage ratio of the radius of the discontinuous surface to the radius of the star (top plot) and radius of the star surface (bottom plot) as functions of the central energy density, where we have considered $\rho_{\rm dis}^+ = 0.8 \times 10^{15}\, \rm g/cm^3$. The left panel corresponds to the equilibrium configurations represented in Fig.~\ref{FifMR1}, while the right panel corresponds to the results of Fig.~\ref{FifMR2}.}
\label{FigRadiusDens}
\end{figure}

To investigate how the size of a NS is influenced by the presence of a CDF core inside it, in Fig.~\ref{FigRadiusDens} we plot the radius (bottom panel) and the ratio $R_{\rm dis}/R$ (top panel) as functions of the central energy density. For fixed $A$ and varying $\alpha$, we see that $R$ decreases with increasing $\alpha$ for very small central densities, but this behavior is reversed after a certain value of $\rho_c$. The ratio $R_{\rm dis}/R$ decreases with increasing $\alpha$. This means that the core size decreases as the energy density jump becomes smaller (i.e., $\alpha$ becomes larger). Meanwhile, for fixed $\alpha$ and varying $A$, we see that $R_{\rm dis}/R$ increases as a consequence of increasing the CDF parameter $A$. According to the upper plot of both panels, the core covers less than $40\%$ of the total radial coordinate of the compact star for very low central densities. However, for central densities above $10^{15}\, \rm g/cm^3$, the core radius is more than $60\%$ of the total radius $R$, indicating that the DE core spans most of the radial coordinate in most stars for the equilibrium configuration families obtained in this study.

\begin{figure}
\begin{center}
\includegraphics[width=6.11cm]{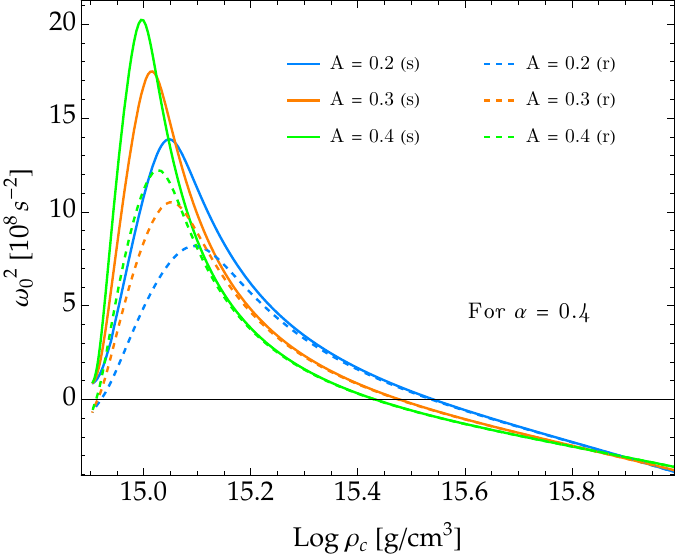}
\includegraphics[width=6.3cm]{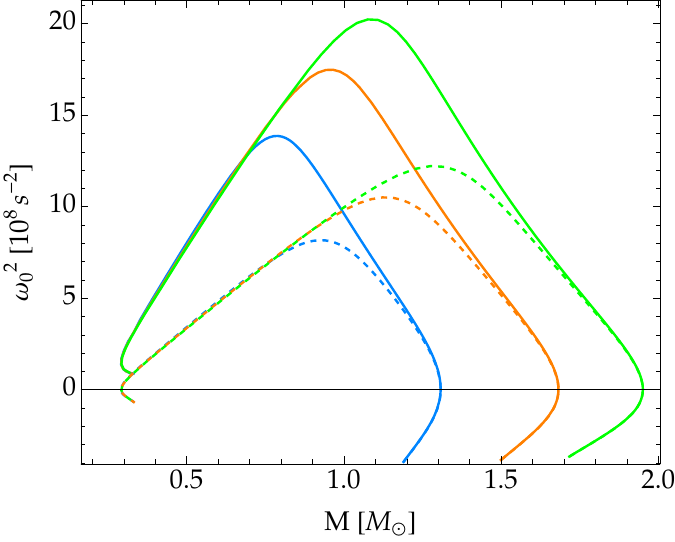}
\end{center}
\caption{Squared frequency of the fundamental vibration mode as a function of the central density (left panel) and of the gravitational mass (right plot) by using $\rho_{\rm dis}^+= 0.8 \times 10^{15}\, \rm g/cm^3$, three values of $A$ and $\alpha= 0.4$ for both slow (solid lines) and rapid (dashed lines) phase transitions. \textit{Source:} Taken from Ref.~\cite{Pretel2024}.}
\label{FifFreq}
\end{figure}

Our next step is to examine the radial stability of the hybrid stellar configurations shown in the $M-R$ diagrams. For a fixed $\alpha= 0.4$ and three values of $A$, figure \ref{FifFreq} illustrates the behavior of the squared frequency of the fundamental mode as a function of central density and mass\footnote{For more details on how to solve the radial vibration equations (\ref{ROEq1}) and (\ref{ROEq2}), and obtain the squared eigenfrequencies, see Ref.~\cite{Pretel2024}.}. According to the left plot, slow phase transitions generate larger $\omega_0^2$ than the rapid phase ones. Note further that decreasing $A$ increases the radial stability of a NS with a dark-energy core in the sense that the critical central density (where $\omega_0^2$ vanishes) becomes increasingly larger with decreasing CDF parameter $A$. From the right plot, we see that the maximum-mass point exactly corresponds to $\omega_0^2= 0$. Notwithstanding, for small masses, where $dM/d\rho_c< 0$ in the mass versus central density relation of Fig.~\ref{FifMR1}, our results show that only rapid phase transitions (see dashed lines in Fig.~\ref{FifFreq}) are able to predict unstable hybrid stars at low central densities, while slow phase transitions are not compatible with the standard stability criterion $dM/d\rho_c >0$.

\begin{figure}
\begin{center}
\includegraphics[width=6.2cm]{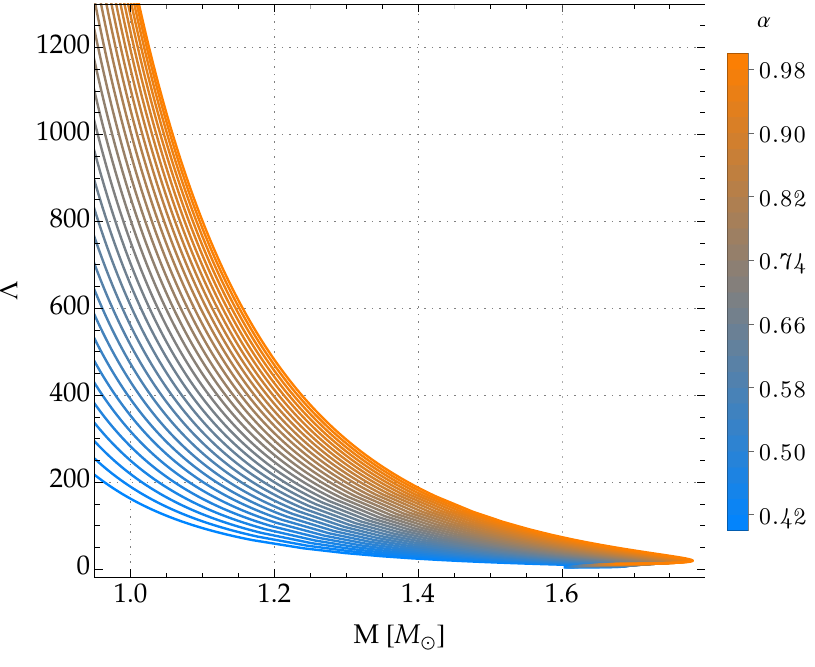}
\includegraphics[width=6.2cm]{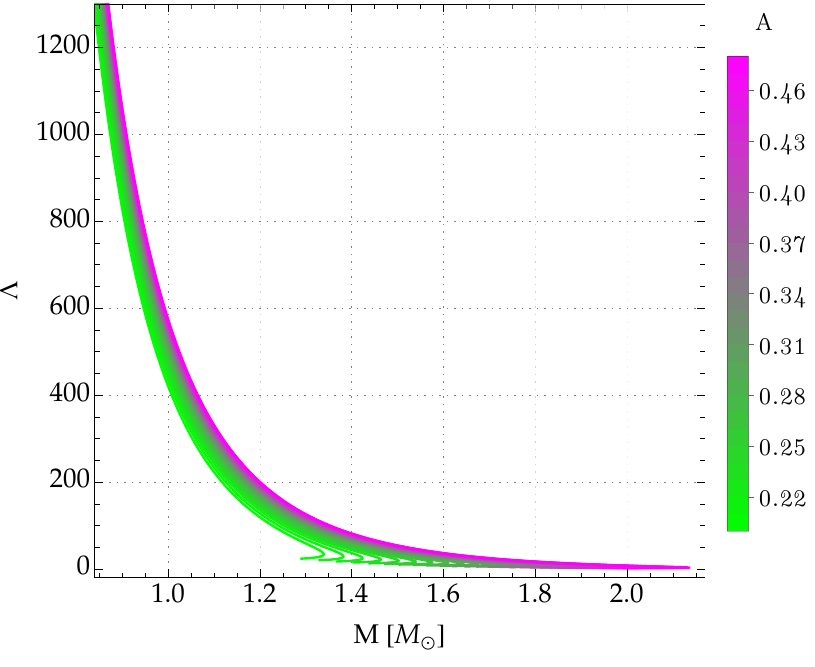}
\end{center}
\caption{Dimensionless tidal deformability $\Lambda$ vs gravitational mass $M$ for the NSs with a CDF core presented in Figs.~\ref{FifMR1} and \ref{FifMR2}. Variations in $\alpha$ have a greater impact on $\Lambda$ than variations in $A$.}
\label{FigLambdaM}
\end{figure}

To calculate the dimensionless tidal deformability $\Lambda$, we must solve Eq.~(\ref{yEq}) taking into account the junction condition at the splitting wall (\ref{BCTidalDef}). The $\Lambda-M$ curves are displayed in Fig.~\ref{FigLambdaM} for the same range of free parameters adopted in Figs.~\ref{FifMR1} and \ref{FifMR2}. It is observed that, given a mass $M$, large tidal deformabilities are produced as a consequence of decreasing the jump in energy density across the interface (that is, as $\alpha$ increases). On the other hand, according to the right plot, the increase in the CDF parameter $A$ implies higher $\Lambda$ for a fixed $M$. Nonetheless, it is important to note that changes in $\alpha$ have a greater effect on the tidal deformability than variations in $A$ even when both ranges have a substantial impact on the mass-radius relations.

\begin{figure}
\begin{center}
\includegraphics[width=6.2cm]{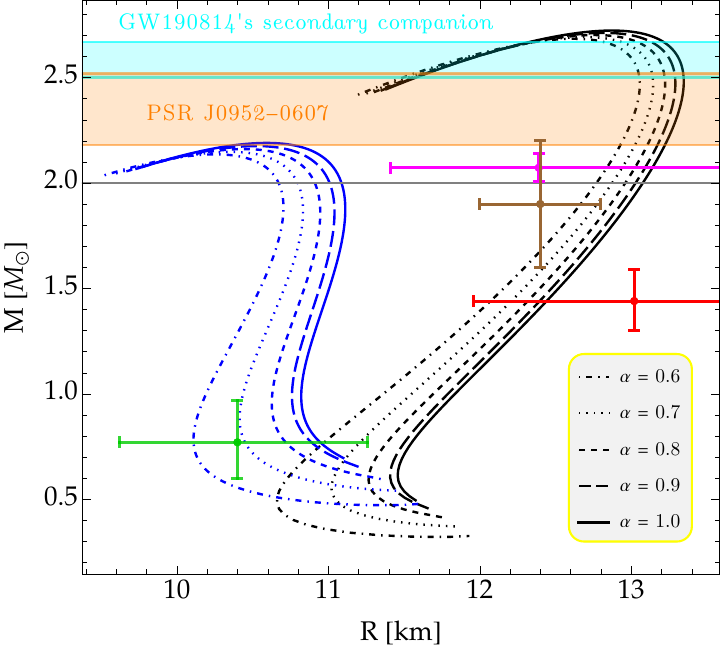}
\includegraphics[width=6.2cm]{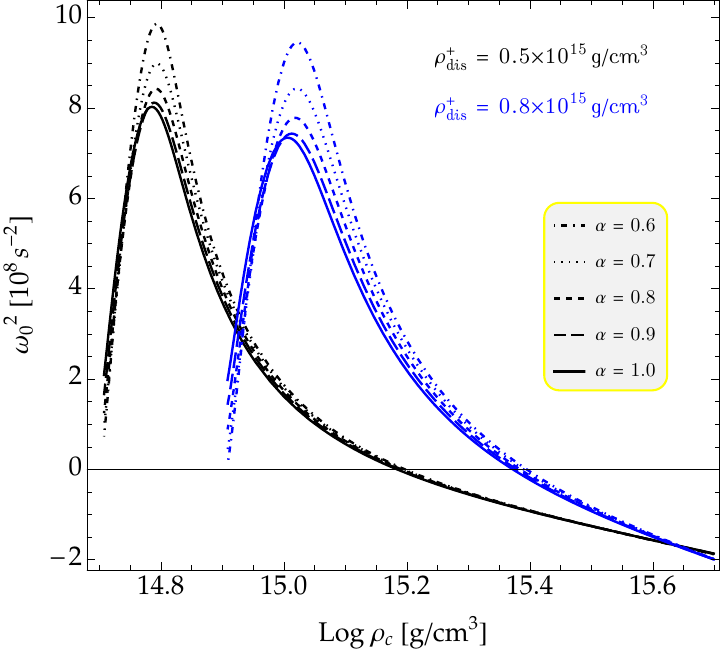}
\end{center}
\caption{Mass-Radius diagram of neutron stars with a dark-energy core (left) and oscillation spectrum under the effect of rapid phase transition (right) for $A= 0.48$ and several values of $\alpha$. Moreover, it has been considered $\rho_{\rm dis}^+= 0.5 \times 10^{15}\, \rm g/cm^3$ (black lines) and $\rho_{\rm dis}^+= 0.8 \times 10^{15}\, \rm g/cm^3$ (blue lines). \textit{Source:} Taken from Ref.~\cite{Pretel2024}.}
\label{FigMRFreqConstr}
\end{figure}

\begin{figure}
\begin{center}
\includegraphics[width=6.24cm]{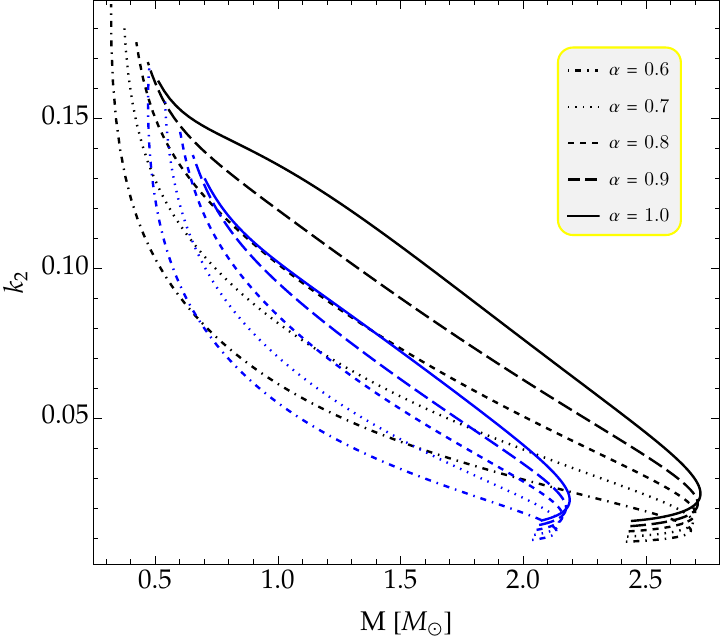}
\includegraphics[width=6.16cm]{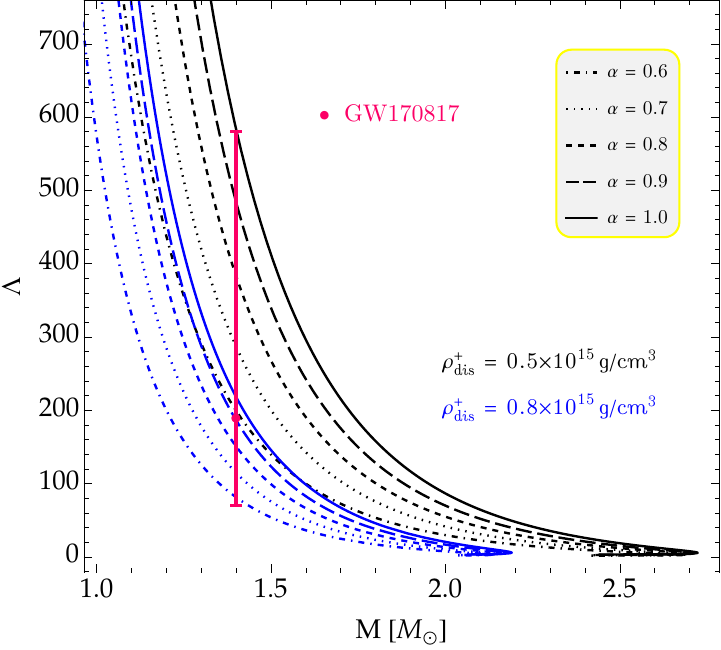}
\end{center}
\caption{Variation of tidal Love number (left) and dimensionless tidal deformability (right) with the gravitational mass of NSs with a CDF core for $A= 0.48$ and two values of $\rho_{\rm dis}^+$ as in Fig.~\ref{FigMRFreqConstr}. The magenta vertical line on the right plot represents the tidal deformability constraint from the GW170817 event, i.e.~$\Lambda_{1.4}= 190_{-120}^{+390}$ \cite{Abbott2018PRL}.}
\label{FigTidalDefConstr}
\end{figure}

Our final task is to examine whether our theoretical predictions are consistent with the astrophysical observations. In that regard, Figs.~\ref{FigMRFreqConstr} and \ref{FigTidalDefConstr} correspond to macroscopic properties of NSs with a CDF core that are in good agreement with different observational measurements, where we have considered two values for the inner energy density $\rho_{\rm dis}^+= 0.5 \times 10^{15}\, \rm g/cm^3$ (black curves) and $\rho_{\rm dis}^+= 0.8 \times 10^{15}\, \rm g/cm^3$ (blue curves). As an additional result, we see that the decrease in $\rho_{\rm dis}^+$ leads to obtaining stars with larger radius and it is possible to get higher maximum masses, and hence generating great compatibility with the secondary component of GW190814 event \cite{Abbott2020AJL}.

According to the left plot of Fig.~\ref{FigMRFreqConstr}, different millisecond pulsars can be very well described with our simple stellar toy model. Moreover, these stars are stable until reaching the maximum mass (when $\omega_0^2 =0$) based on the right hand plot. Compact stars with $\rho_c$ above the critical central density lie in the $\omega_0^2 <0$ region and would collapse into a black hole \cite{KokkotasRuoff2001, Pretel2020MNRAS}. For a star to exist in the Universe, it has to be dynamically stable under radial perturbations. Finally, the right plot of Fig.~\ref{FigTidalDefConstr} reveals that all our results for $\Lambda$ with $\alpha \in [0.6, 1.0]$ are compatible with the tidal deformability constraint from the GW170817 signal \cite{Abbott2018PRL}, i.e.~the first detection of gravitational waves from a binary NS inspiral \cite{Abbott2017PRL}.

\section{Conclusions}\label{Sec4}

In this work we have systematically investigated the effect of a DE core in NSs where the crust is ordinary matter described by a polytropic EoS. Our simple toy model is basically described by the set of parameters $\{\rho_{\rm dis}^+, A, \alpha\}$, where a small $\alpha$ corresponds to a larger jump in the energy density across the phase-splitting surface for a fixed $\rho_{\rm dis}^+$. An increase in $\alpha$ and $A$ lead to increasing the maximum mass values. Furthermore, variations of $\alpha$ over the $M-R$ relations have a larger influence in the low mass region, while the largest impact of the CDF parameter $A$ on the $M-R$ diagram occurs at high masses. With respect to tidal deformability, increasing $\alpha$ and $A$ also leads to increasing $\Lambda$ for a fixed mass $M$. However, increasing $\rho_{\rm dis}^+$ causes smaller tidal deformabilities given a mass $M$.

The most interesting phenomenological cases, compatible with the observational mass-radius measurements and tidal deformability constraints, have been obtained for $\rho_{\rm dis}^+= 0.5 \times 10^{15}\, \rm g/cm^3$ and $\rho_{\rm dis}^+= 0.8 \times 10^{15}\, \rm g/cm^3$ with $A= 0.48$ and a wide range of values for $\alpha$. Our study has therefore shown that the existence of NSs with a DE core is possible in the sense that they are dynamically stable under radial pulsations and are consistent with the recent astrophysical measurements.

Although our stellar model is a kind of toy model, we have seen that it satisfies the observational data. Nevertheless, in future studies it would be convenient to use more realistic EoSs for the hybrid star crust since a large number of EoS models have been proposed by the nuclear physics community. It would also be interesting to examine the inverted case where the core is made of ordinary matter and the crust is composed of DE, as well as the mixed scenario throughout the entire star. The correlation between various global properties such as compactness, moment of inertia and tidal deformability (known as universal relations) is also an issue that needs to be explored in the case of these compact stars. We think that all of these considerations might help us better understand the connection between DE and compact stars.

\section*{Acknowledgments}

JMZP acknowledges support from FAPERJ, Process SEI-260003/000308/2024.


\end{document}